# Towards a Cryogenic CMOS-Memristor Neural Decoder for Quantum Error Correction

Pierre-Antoine Mouny[1,2,3,4], Maher Benhouria[1,2,3], Victor Yon[1,2,3,4], Patrick Dufour[2,3,4], Linxiang Huang[2,3,4], Yann Beilliard[1,2,3,4], Sophie Rochette[1], Dominique Drouin[2,3,4] and Pooya Ronagh[1,5,6,7,*]

[1]*Irréversible Inc., Sherbrooke, Québec, Canada*
{pierre-antoine.mouny | maher.benhouria | victor.yon | sophie.rochette | pooya.ronagh}@irrersible.tech
[2]*Institut Interdisplinaire d'Innovation Technologique (3IT), Université de Sherbrooke, Sherbrooke, Québec, Canada*
{patrick.f.dufour | linxiang.huang | yann.beilliard | dominique.drouin}@usherbrooke.ca
[3]*Laboratoire Nanotechnologies Nanosystèmes (LN2 - IRL 3463) - CNRS, Université de Sherbrooke, Québec, Canada*
[4]*Institut Quantique (IQ), Université de Sherbrooke, Sherbrooke, Québec, Canada*
[5]*Institute of Quantum Computing, University of Waterloo, Waterloo, Ontario, Canada*
[6]*Department of Physics & Astronomy, University of Waterloo, Ontario, Canada*
[7]*Perimeter Institute for Theoretical Physics, Waterloo, Ontario, Canada*

*Abstract*—This paper presents a novel approach utilizing a scalable neural decoder application-specific integrated circuit (ASIC) based on metal oxide memristors in a 180nm CMOS technology. The ASIC architecture employs in-memory computing with memristor crossbars for efficient vector-matrix multiplications (VMM). The ASIC decoder architecture includes an input layer implemented with a VMM and an analog sigmoid activation function, a recurrent layer with analog memory, and an output layer with a VMM and a threshold activation function. Cryogenic characterization of the ASIC is conducted, demonstrating its performance at both room temperature and cryogenic temperatures down to 1.2K. Results indicate stable activation function shapes and pulse responses at cryogenic temperatures. Moreover, power consumption measurements reveal consistent behavior at room and cryogenic temperatures. Overall, this study lays the foundation for developing efficient and scalable neural decoders for quantum error correction in cryogenic environments.

## I. Introduction

In the realm of quantum computing, the fragility of qubits poses a significant challenge. Quantum Error Correction (QEC) protocols offer a solution by encoding information into logical qubits, safeguarding it against errors caused by noise [1]. These protocols periodically generate syndrome signals which characterize errors arising from the qubits. The syndromes are usually processed by a decoder placed at room temperature outside the cryostat which hosts the qubit chip. This approach is not scalable beyond a few hundreds of physical qubits as QEC protocols will produce several gigabytes of syndrome data per second. Routing this amount of information from the qubit chip in the cryostat to the room temperature electronics leads to a wiring bottleneck and is incompatible with the cooling power of current dilution fridges.

Several approaches have been proposed to partially tackle these challenges. References [2][3] present single flux quantum (SFQ)-based architectures. While SFQ is inherently compatible with cryogenic temperatures, current SFQ-based decoders consume too much energy, hindering their scalability. Other works have showcased the development of fast and scalable decoders implementing various types of graph-based decoders including Union-Find [4] and collision clustering [5]. These decoders rely on look-up tables which grow exponentially with the code distance.

On the other hand, neural-network based decoders can perform the decoding task in a single step [6][7]. Notably, recurrent neural networks (RNNs) present an efficient solution for handling the temporal aspects of syndrome processing, thereby reducing data fetching from cryogenic memory blocks [8] (See Fig. 1).

In order to benefit from RNN-based decoder, dedicated cryogenic analog hardware is required. This requires designing a low power analog circuit to be compatible with the cooling power of current dilution fridges (≈1W). This paper presents a small scale neural decoder based on an application specific integrated circuit (ASIC) in 180nm CMOS technology and on metal oxide memristors. The purpose of this design is to demonstrate the viability of a CMOS-memristor decoder but the input and the recurrent layer should be scaled-up in order to correct errors. Section II reports the ASIC architecture including details on the implementation of each layer of the RNN. Section III presents the materials and methods and Section IV shows preliminary experimental results at room temperature and cryogenic temperature.

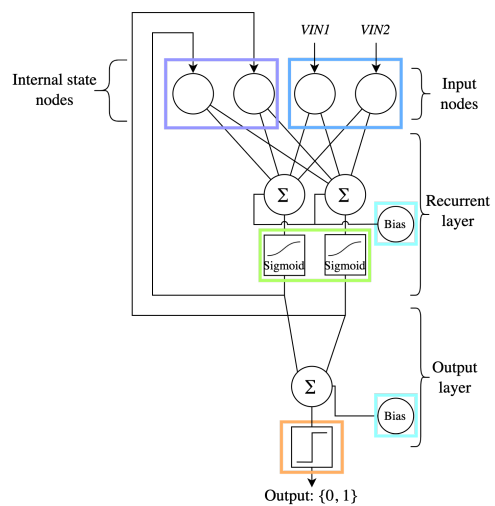

Fig. 1: Small scale recurrent neural network for decoding the Surface code. It comprises a recurrent (input) layer and an output layer. Bias is applied to both layers. The output of the recurrent layer is routed back to be used as new input to the internal states' nodes along with a new voltage vector.

## II. ASIC Architecture

The recurrent neural network is based on in-memory computing with memristor crossbars. Memristors are non-volatile memories which exhibit fully analog resistance programming and cryogenic compatibility down to 4.2K [9]. Thus, their conductance can be arbitrarily programmed using voltage pulses (see Fig. 2) to encode the neural weights of a given neural network [10].

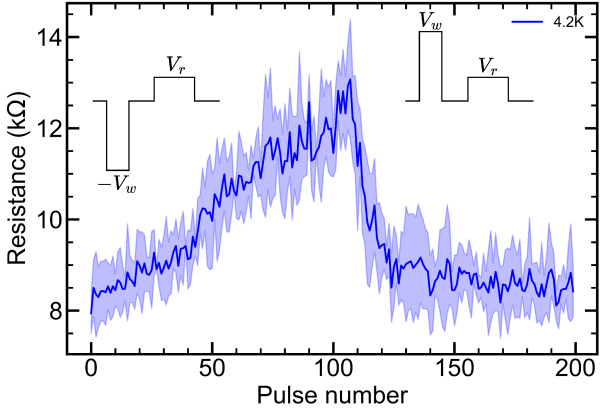

Fig. 2: Pulse programming of a TiO$_x$ memristor at 4.2K. 100 negative pulse trains are applied followed by 100 positive pulse trains. The variability depicts the standard deviation arising from 10 measurements.

By applying a voltage pulse at the VIN1 and VIN2 inputs, a vector matrix multiplication is performed. Using Kirchhoff's current law and Ohm's law, the output current is given by: $I_j^\pm = \sum_i G_{ij}^\pm V_i^{IN}$. These operations correspond to the forward pass of a neural network, where the input pulses encode the input data (syndromes in the case of a decoder).

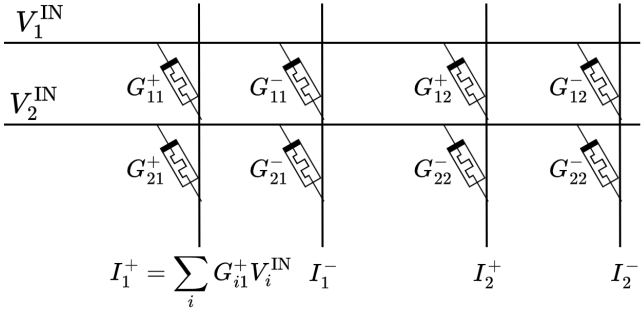

Fig. 3: Vector matrix multiplication with a memristor crossbar. A pair of memristors is used to encode a single weight. The current at the output of each line represents either the positive or negative part of an element of the matrix-vector multiplication.

The ASIC performs the readout of this output current and serves as a proof of concept for the complete decoder. It can be decomposed into three parts, as shown in Fig. 4: the first and second parts form the input and the recurrence layers of the neural network, decomposed as two distinct layers on the ASIC decoder. The third part is the output layer.

### A. Input Layer

The input layer consists of two channels connected to the first part of the memristor crossbar. VIN1 and VIN2 are binarized inputs i.e, the input is either a 0V input pulse or a 1.8V input pulse corresponding respectively to a logical 0 and a logical 1. These input pulses pass through the input weights resulting in the positive and negative current components of the vector matrix multiplication.

A current buffer serves as an interface to measure these currents. It includes a subtractor circuit that produces a current equivalent to the difference between the positive and negative inputs. Depending on the chosen weights, the output current can be positive or negative. The circuit also applies a bias voltage to ensure that the outputs of the memristor crossbar share the same voltage (1.2V), preventing sneak paths. The output current, depending on the memristor conductances, is fed to a resistive circuit implementing a sigmoid activation function [11].

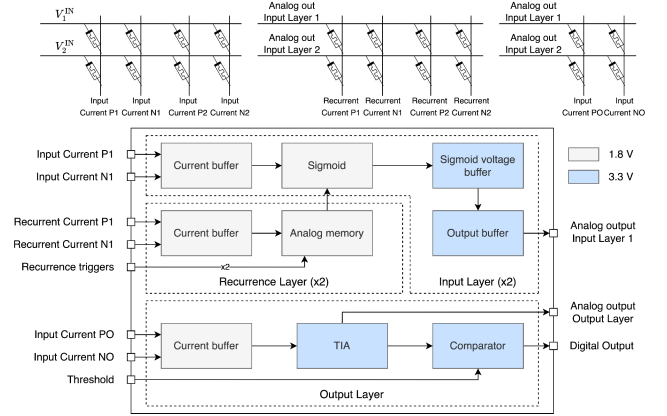

Fig. 4: Decoder ASIC block diagram. The output currents of the two memristor crossbars are fed to the input current P and N, and recurrent current P and N pins of the ASIC decoder chip. The input layer and recurrence layer circuits depicted on this figure are placed twice on the ASIC decoder chip yielding to two channels with subsequent naming e.g., Input current P1, Sigmoid_1 and Sigmoid_2.

Since the sigmoid output ranges from 0 to 1.8V, a voltage buffer with a gain of ⅓ is used to adjust the output voltage, allowing it to span from 1.2V to 1.8V. This adjustment guarantees that the output of the first layer falls within a range compatible with the operational limits of the memristors.

### B. Recurrence Layer

The recurrence layer interfaces with a second memristor crossbar, receiving input voltages from the outputs of the previous layers. During the same computing step, these voltages pass through the recurrent weights and the current buffers as depicted in Fig. 4, effectively pre-computing the recurrent currents.

This recurrent current is stored in an analog memory. The analog current memory [12] enables the storage of this current value, which can be either positive or negative, and subsequently releases it after a specified delay. The storage process involves maintaining the gate voltage of a transistor, which functions as a current source, in a capacitor. The circuit's operation is controlled by two triggers: the first trigger enables the storage of the current value (latch), while the second trigger initiates the release process. The recurrent current is fed into the sigmoid circuit, where it is summed with the input layer current during the next computing step.

### C. Output Layer

The output layer adopts the same architecture as the first layer, but with a binary step activation function. A transimpedance amplifier (TIA) is implemented to transform the current output from the current buffer into a voltage, facilitating external readout of the analog signal. A comparator with an external threshold performs the binary classification, allowing the interpretation of the neural network output as a digital signal.

### III. MATERIALS AND METHODS

The ASIC was fabricated using the TSMC 180nm CMOS process, with a die size of 1 x 2 mm, and then wirebonded to a Dual-In-Line (DIP) chip carrier. A custom PCB test platform was designed to characterize its performance. This main PCB also includes decoupling capacitors, a set of connectors for a memristors chip carrier, and another set for providing external power supplies and interfacing with the IOs of the ASIC. No other components were integrated on this PCB, as it is specifically designed for cryogenic tests where only the ASIC and the resistor crossbar undergo cooling. For the following electrical characterizations, memristors were replaced with fixed discrete resistors. This allows to validate the behavior of the CMOS blocks and enables the evaluation of the overall performance of the cryogenic neural decoder system independently.

The next section provides details on the characterization tests, which were conducted in two steps: A) validation of circuits at room temperature and B) cryogenic tests.

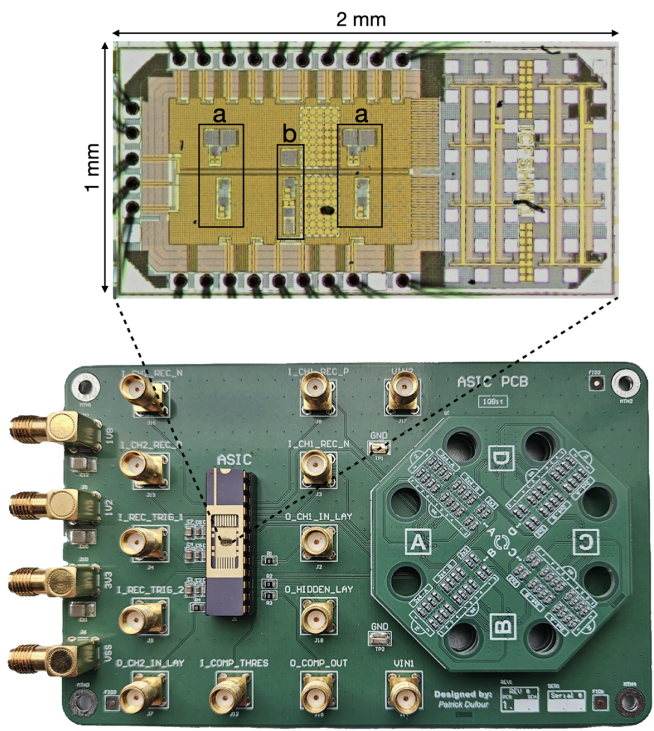

Fig. 5: Main test PCB with the 180nm CMOS decoder chip - (a) Input layer with recurrence; (b) Output layer. The decoder chip is wirebonded on a DIP sample holder while the synaptic weights are soldered on a custom chip carrier.

#### A. Room Temperature Validation

These tests aim to validate the performance of the ASIC and compare it with simulation results before initiating cryogenic tests, where the outcomes may deviate from the simulations.

The shape of the sigmoid activation function is validated through DC tests. The ASIC is powered with a benchtop power supply providing 3.3V, 1.8V and 1.2V, while two DC currents are applied to the input layer. These currents are generated by a resistance placed between each input pad and 1.8V. The current value is then varied by sweeping the resistance value :

$$Input\ Current\ = \frac{1.8\,V - 1.2\,V}{R}$$

The dynamic behavior is also validated with different input configurations. Voltage pulses of 1μs are applied to VIN1 and VIN2 using a LOTUS characterization platform from Advanced MicroTesting which provides 32 arbitrary pulse measurement units. Recurrence control signals are also triggered to ensure the proper storage and release of currents. Pulse shapes at the output of each neuron of the input layer, as well as the analog output of the second layer, are then observed.

#### B. Cryogenic Tests

The main PCB is placed at the 1K-stage of an ICEoxford DRY ICE 1.5K cryostat for the cryogenic tests. The ASIC is supplied with 1.2V, 1.8V and 3.3V while being cooled down. The shape of the sigmoid activation function is investigated through current sweeps applied to the input of the sigmoid circuit. This is achieved by sweeping the gate voltage of the compliance transistors of the LOTUS characterization platform. However, the dynamic behavior study is performed with a single set of input (VIN1=VIN2=2.5V) and the same resistance configuration as room temperature.

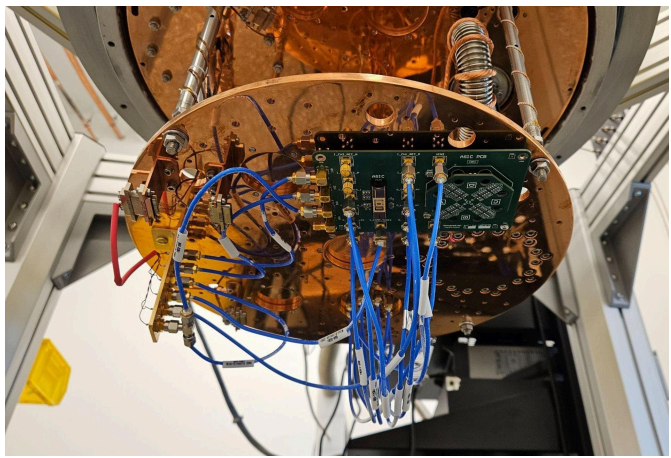

Fig. 6: Cryogenic temperature tests setup. The main PCB is placed in an ICEoxford DRY ICE cryostat and cooled down to 1.2K.

The same set of tests is performed at different temperatures including the base temperature of the cryostat 1.2K, 4.2K, 35K and 77K. Only the temperature of the 1K-stage of the cryostat is adjusted while the 4K-pot and 50K-stage are kept at 3.2K respectively 60K.

### IV. RESULTS

#### A. Room Temperature Validation

Fig. 7 presents the sigmoid shape from channel 2 of one sample of the decoder ASIC, comparing it to the sigmoid from simulations done with Cadence Spectre. Measurements were limited to the expected operating range of [-500uA, to 500uA].

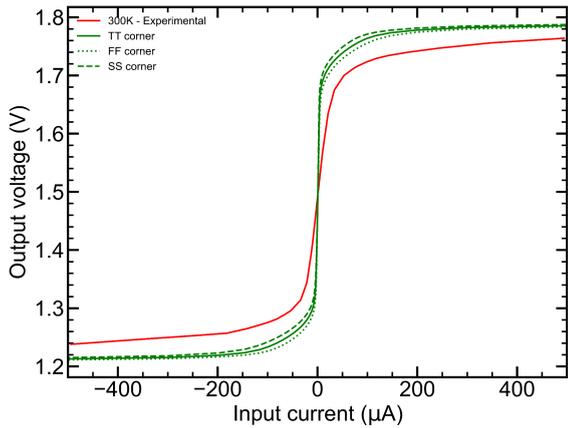

Fig. 7: Analog sigmoid activation function. The green curves depict the circuit simulation results at different process corners: typical-typical (TT) and the two even corners fast-fast (FF) and slow-slow (SS). For example, fast-fast corners exhibit carrier mobilities that are higher than normal for NMOS and PMOS in the circuit.

Fig. 8 presents pulses obtained from different input configurations with two consecutive input pulses of 1us/1.8V corresponding to 2 rounds of computation. The second pulses depicted in the 'Out 11' and 'Out 01' plots show the effect of recurrence.

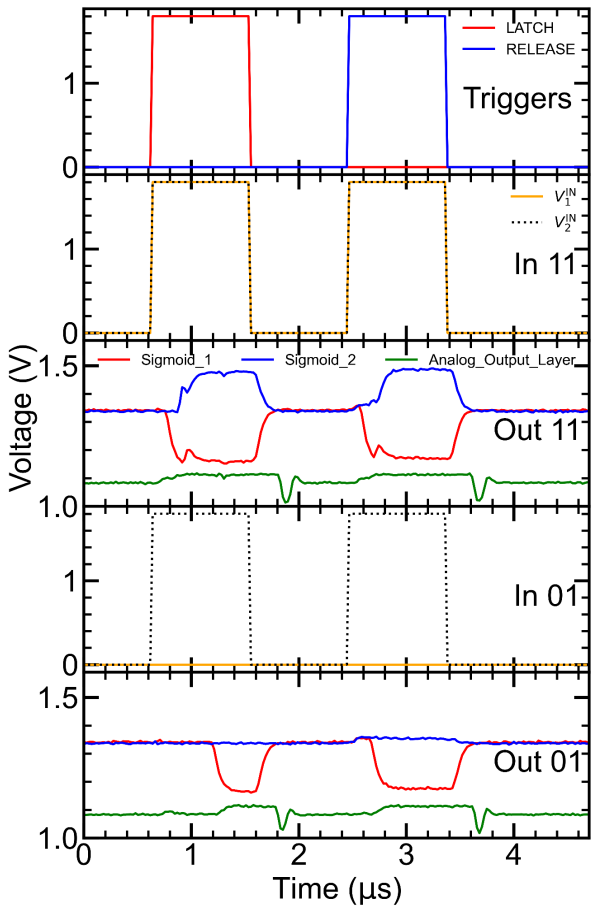

Fig. 8: Pulse response of the ASIC decoder for two different sets of input pulses at room temperature.

## B. Cryogenic Tests

The shape of the sigmoid activation function from the DC sweep test is shown in Fig. 9. The measurements are performed at various cryogenic temperatures to compare the impact of temperature on the sigmoid circuit behavior.

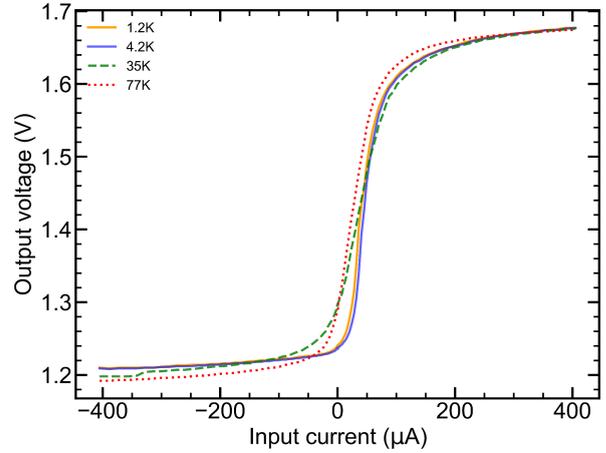

Fig. 9: Analog sigmoid activation function at different cryogenic temperatures. Current was directly injected at the Input Current P1 and Input Current N1 of the Sigmoid circuit using a current pulse generator.

Fig. 10 shows the output pulses of the input layer and output layer at various cryogenic temperatures for a 1us/2.5V input pulse applied at both VIN1 and VIN2.

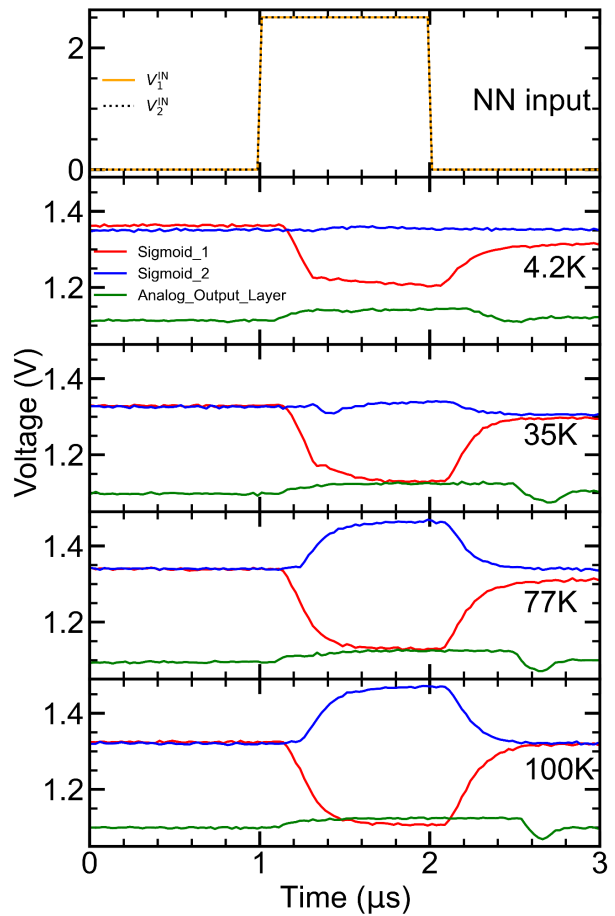

Fig. 10: ASIC decoder pulse response at various cryogenic temperatures.

Power consumption measurements at both room and cryogenic temperatures are summarized in Table I.

| T (K) | Power consumption | | |
|---|---|---|---|
| | 1.8V supply | 3.3V supply | Total |
| 1.2K | 3.2mW | 10.2mW | 13.4mW |
| 4.2K | 3.3mW | 10.9mW | 14.2mW |
| 35K | 3.4mW | 10.9mW | 14.3mW |
| 300K | 3.4mW | 11.9mW | 15.3mW |

TABLE I. POWER CONSUMPTION OF THE ASIC WITH RESPECT TO THE TEMPERATURE AND THE SUPPLY VOLTAGES.

## V. DISCUSSION

Measurements at room temperature show that the chip's performance align well with simulation results. The DC sigmoid shape corresponds to a nonlinear activation function, but with a slight offset from simulated values due to IC manufacturing process variations. However, these variations can be effectively mitigated through the implementation of hardware-aware techniques during the neural network training [8].

The pulse response of the IC also matches the anticipated results. Slightly longer rise and fall times were noticed, and these can be attributed to the presence of large capacitive loads introduced by the measurement circuits and cables routing the signal.

Preliminary cryogenic characterizations of the analog sigmoid indicate that the shape of the activation function depends on the temperature. At lower cryogenic temperatures (1.2K and 4.2K), the transition between the asymptote and linear regime is sharper and happens at lower absolute input currents. This results in a larger slope in the linear regime which is a closer behavior to the sigmoid circuit simulation reported in Fig. 7. As the temperature increases (35K and above), the transition between the asymptotic regime and the linear regime is smoothed and the linear regime slope is reduced similar to the 300K sigmoid behavior. The shape of this sigmoidal activation function is dependent on the threshold voltage of the transistors [11] which commonly increases as the temperature decreases [13] which support the shape mismatch exhibited by our sigmoid circuit implementation.

This sigmoid behavior discrepancy is also observed during pulse characterizations. At lower cryogenic temperatures, the output pulse amplitude from the sigmoid activation function is lower for both channels leading to no output pulse for Sigmoid_2 at 4.2K and only a slight kink at 35K. However the neural network output is unchanged with temperature which indicates that, in this particular configuration, the inference result will be identical for all temperatures despite different sigmoid outputs. The rise time is unaffected by the temperature while the fall time exhibited at 4.2K is larger than at other temperatures. A delay of 100ns is exhibited by the output pulses at all temperatures which is partly due to the parasitic capacitances introduced by the cabling of the cryostat.

Finally, the ASIC decoder shows a constant power consumption for its 1.8V supply while its power consumption for the 3.3V supply increases with temperature. The overall power consumption is not compatible with a substantial scaling of this architecture to correct large Surface codes. However, the power consumption could be drastically reduced by using lower CMOS technological nodes (e.g., TSMC 28nm or GlobalFoundries 22nm FDX).

## VI. CONCLUSION

In conclusion, this study introduces a novel ASIC architecture utilizing metal oxide memristors for efficient neural decoding in Quantum Error Correction (QEC) applications. Cryogenic characterization demonstrates stable performance down to 1.2K, indicating potential scalability and reliability in cryogenic environments. While challenges exist, such as sigmoid behavior discrepancies at lower temperatures, this research lays a solid foundation for the development of robust neural decoders which is a crucial step for scaling up quantum computers and enabling key industrial applications.

ACKNOWLEDGEMENTS

We thank our editor, Marko Bucyk, for his careful review and editing of the manuscript. The authors acknowledge the financial support received through the NSF's CIM Expeditions award (CCF-1918549). P. R. acknowledges the financial support of Mike and Ophelia Lazaridis, Innovation, Science and Economic Development Canada (ISED), and the Perimeter Institute for Theoretical Physics. Research at the Perimeter Institute is supported in part by the Government of Canada through ISED and by the Province of Ontario through the Ministry of Colleges and Universities. This work was supported by the Natural Sciences and Engineering Research Council of Canada (NSERC). We acknowledge Christian Lupien and the Institut Quantique for their assistance with electrical characterization at cryogenic temperatures. We thank Pr. Fabien Alibart (Université de Sherbrooke) for his comments on the manuscript. LN2 is a French-Canadian joint International Research Laboratory (IRL-3463) funded and co-operated by the Centre national de la recherche scientifique (CNRS), the Université de Sherbrooke, the Université de Grenoble Alpes (UGA), the Ecole centrale de Lyon (ECL), and the Institut national des sciences appliquées de Lyon (INSA Lyon). It is supported by the Fonds de recherche du Québec - Nature et technologie (FRQNT). We would like to acknowledge CMC Microsystems for the provision of products and services that facilitated this research, including CAD tools and fabrication services using the 180-nanometre CMOS technology from TSMC.